%% file: conference_101719.tex
\documentclass[10pt,conference]{IEEEtran}
\IEEEoverridecommandlockouts

\usepackage{cite}
\usepackage{url}
\usepackage{glossaries}
\usepackage{amssymb,amsfonts}
\usepackage{amsmath}
\usepackage{algorithm}
\usepackage{algpseudocode}
\usepackage{graphicx}
\usepackage{textcomp}
\usepackage{xcolor}
\usepackage{fancyhdr}
\usepackage{subfigure}
\usepackage[top=0.75in, bottom=1.0in, left=0.625in, right=0.625in]{geometry}

\usepackage{ifthen}

\def\BibTeX{{\rm B\kern-.05em{\sc i\kern-.025em b}\kern-.08em
    T\kern-.1667em\lower.7ex\hbox{E}\kern-.125emX}}

\newboolean{showcomments}
\setboolean{showcomments}{true} 

\ifthenelse{\boolean{showcomments}}{%
  \newcommand{\mynote}[2]{%
    \fbox{\bfseries\sffamily\scriptsize#1}%
    {\small$\blacktriangleright$%
    \textsf{\textcolor{red}{{\em #2}\bf}}%
    $\blacktriangleleft$}%
  }%
  \newcommand{\notered}[1]{\textcolor{red}{[{\bf #1}]}}%
  \newcommand{\notegreen}[1]{\textcolor{green}{[{\bf #1}]}}%
}{%
  \newcommand{\mynote}[2]{}%
  \newcommand{\notered}[1]{}%
  \newcommand{\notegreen}[1]{}%
}

\newcommand*{\affmark}[1][*]{\textsuperscript{#1}}

\input{glossary}

\begin{document}

\title{\vspace{-4mm} Hierarchical Decision Mamba Meets Agentic AI: \\ A Novel Approach for RAN Slicing in 6G \\ \vspace{-4mm}}
\vspace{-5pt}

\author{\IEEEauthorblockN{Md~Arafat~Habib\affmark[1], Medhat Elsayed\affmark[2], Majid Bavand\affmark[2], Pedro Enrique Iturria Rivera\affmark[2], \\ Yigit Ozcan\affmark[2], and  Melike Erol-Kantarci\affmark[1], \IEEEmembership{
Fellow,~IEEE}}
\IEEEauthorblockA{\affmark[1]\textit{School of Electrical Engineering and Computer Science, University of Ottawa, Canada}}\affmark[2]\textit{Ericsson Inc., Ottawa, Canada}\\
Emails:\{mhabi050, melike.erolkantarci\}@uottawa.ca,\\ \{medhat.elsayed, majid.bavand, pedro.iturria.rivera, yigit.ozcan\}@ericsson.com \vspace{-1em}
\vspace{-5pt}
}

\maketitle

\thispagestyle{fancy}   
\fancyhead{}                
\cfoot{}
\renewcommand{\headrulewidth}{0pt} 

\begin{abstract}
Radio Access Network (RAN) slicing enables multiple logical networks to exist on top of the same physical infrastructure by allocating resources to distinct service groups, where radio resource scheduling plays a key role in ensuring compliance with slice-specific Service-Level Agreements (SLAs). Existing configuration-based or intent-driven Reinforcement Learning (RL) approaches usually rely on static mappings and SLA conversions. The current literature does not integrate natural language understanding with coordinated decision-making. To address these limitations, we propose an Agentic AI framework for 6G RAN slicing, driven by a super agent built using Hierarchical Decision Mamba (HDM) controllers and a Large Language Model (LLM). The super agent interprets operator intents and translates them into actionable goals using the LLM, which are used by HDM to coordinate inter-slice, intra-slice, and self-healing agents. Compared to transformer-based and reward-driven baselines, the proposed Agentic AI framework demonstrates consistent improvements across key performance indicators, including higher throughput, improved cell-edge performance, and reduced latency across different slices.
\end{abstract}


\vspace{-5pt}

\section{Introduction}  
\label{s1}

\glspl{RRS} in the \gls{RAN} allocate resources according to slice-specific \glspl{SLA}. Inter-slice scheduling determines how resources are divided among slices, while intra-slice scheduling maps those resources to \glspl{UE} based on traffic and \gls{QoS} requirements \cite{4}. In 6G networks, the \gls{RRS} must be aware of the diverse needs of different applications. Recent studies have proposed building \glspl{RRS} using structured intents that include predefined mappings to \gls{SLA} parameters \cite{2,4}. However, these works rely on static \gls{RL} policies with fixed slice weights, and they lack near real-time adaptivity or self-corrective control. Moreover, their RL architectures incur high control latency and lack contextual grounding from external knowledge.  

Considering these limitations of the previous works, we propose an Agentic AI framework guided by \gls{HDM} \cite{7} for intelligent and adaptive \gls{RAN} slicing in 6G networks. The proposed framework introduces multi-level autonomy using a super agent that interprets operator intents, coordinates specialized agents, and enables dynamic policy selection within a hierarchical control structure. The key contributions of this work are as follows:  
\begin{itemize}
    \item To the best of our knowledge, for the first time in the literature, a \gls{HDM} architecture for Agentic AI-based network management is presented. This low-latency control framework utilizes state-space sequence modeling to achieve faster and more efficient decision processing.
    \item We propose a self-corrective control mechanism to perform self-healing that continuously adjusts slice weights and priorities to mitigate performance deviations and ensure consistent \gls{SLA} compliance.
    \item A Hybrid \gls{RAG}-based decision-making framework is introduced, integrating static knowledge sources (e.g., 3GPP and O-\gls{RAN} standards) with dynamic network analytics to enable context-aware and explainable decisions.
   \item To the best of our knowledge, we present the first coordinated Agentic AI framework that jointly orchestrates inter-slice provisioning, intra-slice scheduling, and self-healing for adaptive \gls{RAN} management.
\end{itemize}
\vspace{-6pt}

\section{System Model}
\label{s2}
\vspace{-6pt}

\subsection{Network Model}
We consider a millimeter-wave (mmWave) \gls{MIMO} system composed of $c = \{1, 2, \ldots, C\}$ cells. Each \gls{BS} is equipped with $N_t$ transmit antennas and serves $u = \{1, 2, \ldots, U\}$ \glspl{UE} distributed across $s = \{1, 2, \ldots, S\}$ active \gls{RAN} slices. Each slice $s$ is associated with specific \gls{QoS} requirements such as throughput, latency, or reliability, according to its \gls{SLA}. The system operates over a total bandwidth $B$, divided into $R$ \glspl{RBG}, which form the minimum allocation unit for the radio resource scheduling. The minimum scheduling time interval is a \gls{TTI} of duration $T_{\text{TTI}}$. The absolute time at global scheduling step $n$ is $t_n = n\,T_{\text{TTI}}$, where $n$ indexes the decision epochs of all controllers. The inter-slice \gls{RRS} allocates \glspl{RBG} among slices, while intra-slice \gls{RRS} distributes them among \glspl{UE} within a slice. The proposed scenario uses \gls{TDD} and assumes perfect channel estimation via pilot transmission. \gls{SE} per \gls{UE} varies with time but remains constant across \glspl{RBG} within the same \gls{TTI}.

We consider a seven-cell hexagonal layout where inter-cell and inter-sector interferences are treated as noise. The spectral efficiency for \gls{UE} $u$ connected to \gls{BS} $b$ is expressed as:
\begin{equation}
\text{SE}_{b,u}(n) = 
\log_2 \!\left( 1 + \frac{\text{RSRP}_{b,u}}{I^{\text{inter}}_{b,u} + \sigma^2} \right).
\label{eq:se_mimo}
\end{equation}
Here, $\sigma^2$ denotes the noise power and $I^{\text{inter}}_{b,u}$ represents the inter-cell interference computed from the six strongest neighboring \glspl{BS}. The term $\text{RSRP}_{b,u}$ denotes the reference signal received power between \gls{BS} $b$ and \gls{UE} $u$. At each step $n$, the radio resource scheduler allocates $R_s(n)$ \glspl{RBG} to each slice $s$, ensuring that
\begin{equation}
\sum_{s=1}^{S} R_s(n) = R, \quad
r_u(n) = \frac{R_s^u(n)}{R} \ast B \ast \text{SE}_{b,u}(n),
\label{eq:rbg_throughput_combined}
\end{equation}
where $R$ is the total number of available \glspl{RBG}, and $r_u(n)$ denotes the instantaneous served throughput for \gls{UE} $u$ in slice $s$, computed based on its allocated RBGs $R_s^u(n)$, system bandwidth $B$, and spectral efficiency $\text{SE}_{b,u}(n)$. The effective throughput, considering data availability in the buffer $b_u(n)$, is: $r_u^{\text{eff}}(n) = \min(r_u(n), b_u(n))$.

The average latency experienced by \gls{UE} $u$ at step $n$ is estimated as $l_u(n) = \frac{b_u(n)}{r_u^{\text{eff}}(n)}$, where $b_u(n)$ is the instantaneous buffer occupancy. Lower latency values are expected for \gls{URLLC} slices due to reduced queueing and higher service rates.

To capture long-term user performance, the average served throughput and the fifth-percentile served throughput are defined as
\begin{equation}
g_u(n) = \frac{1}{m} \sum_{i=n-m+1}^{n} r_u(i),
\label{eq:longterm_throughput}
\end{equation}
\begin{equation}
f_u(n) = P_{5\%}\big(r_u(n-m+1), \ldots, r_u(n)\big),
\label{eq:5th_percentile}
\end{equation}
where $P_{5\%}(\cdot)$ computes the fifth-percentile value of the served throughput samples. These metrics are used to evaluate \gls{URLLC} latency compliance, \gls{eMBB} throughput, and cell-edge user experience.

\subsection{Slice Types}

\begin{figure}[!t]
\centerline{\includegraphics[width=1\linewidth]{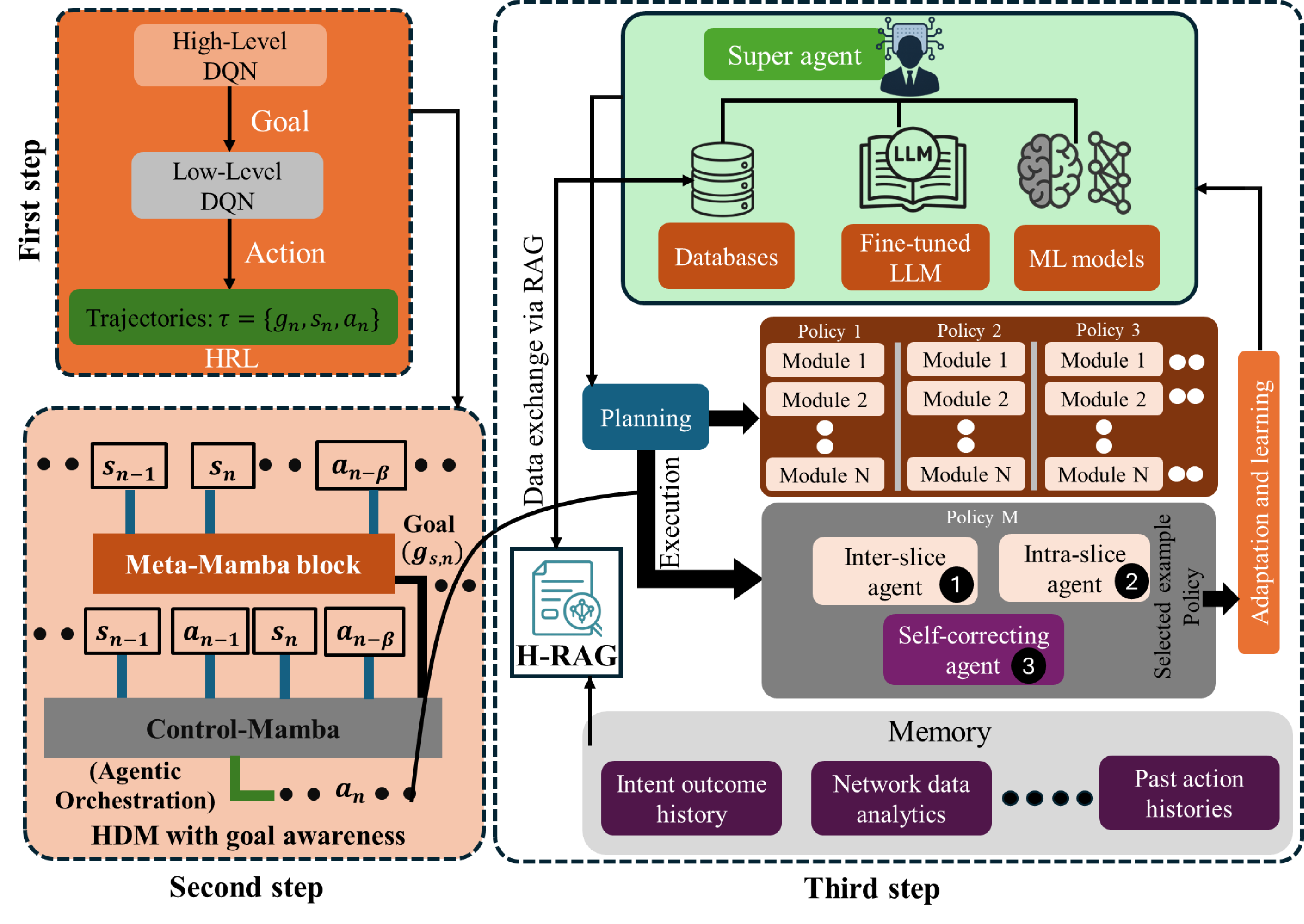}}
\caption{Agentic architecture of the proposed system guided by Hierarchical Decision Mamba}
\label{fig1}
\vspace{-1em}
\end{figure}

We consider three slice types: \gls{eMBB}, \gls{URLLC}, and \gls{BE} \cite{2}. Each slice is associated with a distinct set of \gls{SLA} requirements that guide the \glspl{RRS}.

\textbf{1) \gls{eMBB} Slice:} \gls{eMBB} \glspl{UE} require high throughput with relaxed latency and reliability constraints, typical of applications such as ultra-HD video streaming. The key \gls{QoS} parameters are: (i) average served throughput $r_{\text{embb}}(n) \ge r^{\text{req}}_{\text{embb}}$, (ii) average latency $\ell_{\text{embb}}(n) \le \ell^{\text{req}}_{\text{embb}}$, and (iii) packet loss rate $p_{\text{embb}}(n) \le p^{\text{req}}_{\text{embb}}$.

\textbf{2) URLLC Slice:} The QoS parameters are similar to eMBB but with stricter thresholds: $r_{\text{urllc}}(n) \ge r^{\text{req}}_{\text{urllc}}$, $\ell_{\text{urllc}}(n) \le \ell^{\text{req}}_{\text{urllc}}$, and $p_{\text{urllc}}(n) \le p^{\text{req}}_{\text{urllc}}$. 

\textbf{3) BE Slice:} BE traffic represents background or delay-tolerant applications such as social media or non-real-time multimedia. The main intents are to maintain the long-term served throughput $g_{\text{be}}(n) \ge g^{\text{req}}_{\text{be}}$ and the fifth-percentile throughput $f_{\text{be}}(n) \ge f^{\text{req}}_{\text{be}}$. BE UEs are activated or deactivated every $n_{\text{be}}$ steps with equal probability.
 
\subsection{Problem Formulation}

The objective of the proposed Agentic AI framework is to derive a structured hierarchical policy that guides the decision-making of the super-agent under dynamically evolving \gls{RAN} conditions. The super-agent policy $\pi_S$ must maximize the cumulative network utility while satisfying the \gls{SLA} requirements of all active slices. Formally:
\begin{equation}
\begin{aligned}
\pi_\phi^{*}
&=\arg\max_{\pi_\phi}\;
\mathbb{E}_{\tau \sim \pi_\phi}
\!\left[\sum_{n=1}^{N} \mathcal{R}(s_n,a_n)\right],\\[2pt]
\text{s.t.}\quad 
&\mathrm{S}_{\mathrm{QoS}}(s_n)\in\mathcal{C}_g,\ \forall n,\\[2pt]
&\boldsymbol{\theta}_n^{(\mathrm{RBG})}\in\Theta_{\mathrm{A_{ID}}(n)},\ \forall n,\\[2pt]
&\sum_{s=1}^{S} \theta_{n,s}^{(\mathrm{RBG})} \le R_{\max}.
\end{aligned}
\label{eq:pf}
\end{equation}

In \eqref{eq:pf}, $\pi_\phi$ denotes the goal-conditioned policy parameterized by $\phi$, and $\pi_\phi^{*}$ represents the optimal HDM policy. The expectation $\mathbb{E}_{\tau \sim \pi_\phi}[\cdot]$ is over trajectories $\tau = (s_0, a_0, \ldots, s_N)$ generated by the transition dynamics $\mathcal{T}$. At each time step $n$, the super-agent observes the system state $s_n$, selects an orchestration action $a_n = \pi_\phi(s_n)$ based on the intent-derived goal $g_n$, and receives a reward $\mathcal{R}(s_n, a_n)$ that reflects progress toward the goal region $\mathcal{C}_g$. The constraint $\mathrm{S}_{\mathrm{QoS}}(s_n)\in\mathcal{C}_g$ enforces slice-specific \gls{QoS} compliance. The vector $\boldsymbol{\theta}_n^{(\mathrm{RBG})} \in \Theta_{\mathrm{A_{ID}}(n)}$ limits the action to the feasible RBG-allocation set of the selected agent, while $\sum_{s=1}^{S} \theta_{n,s}^{(\mathrm{RBG})} \le R_{\max}$ ensures that the total allocated RBGs do not exceed the available system budget.

\section{Proposed Methodology Based on Agentic AI}

In this work, we propose a complete Agentic AI system for \gls{RAN} slicing comprising four different agents. The Agentic AI architecture with the super agent, inter-slice, intra-slice, and self-healing agents, is illustrated in Fig. \ref{fig1}.

\textbf{1) Super agent powered by \gls{LLM} and Decision Mamba:} The super agent conducts high-level planning, coordination, and orchestration by managing goals, resolving conflicts, and aligning the actions of subordinate agents toward a unified objective. It is implemented using a fine-tuned Llama~3.2 model as described in~\cite{6}, interfacing directly with network operators to interpret intents while utilizing a \gls{H-RAG} framework for contextual grounding. The \gls{H-RAG} integrates static knowledge $D_s$ (e.g., 3GPP standards and policy rules) and dynamic knowledge $D_d(n)$ (real-time \glspl{KPI} and slice performance data). Given an input intent $x$, a query encoder $f_{\text{enc}}$ produces an embedding $z=f_{\text{enc}}(x)$, and relevant context $\mathcal{C}=\mathcal{C}_s \cup \mathcal{C}_d$ is retrieved from both sources to enrich the intent representation. The resulting enriched embedding supports goal formulation and adaptive orchestration, enabling coherent, context-aware collaboration across the agent hierarchy without replacing local decision modules.

Natural-language intents (e.g., ``\gls{URLLC} latency $\le 2~ms$'' or ``increase \gls{eMBB} throughput by $10\%$'') are processed by an \gls{LLM} to extract structured slice-specific \gls{QoS} targets. These targets directly guide the Agentic AI orchestration decisions. While the \gls{LLM}-based super agent enables semantic intent interpretation, an additional layer of intelligence is required to translate high-level goals into concrete orchestration actions. To this end, we integrate a Decision Mamba-based control algorithm that allows the super agent to intelligently select and parameterize the required functional agents in the \gls{RAN} slicing environment. The orchestration process is modeled as a goal-conditioned semi \gls{MDP}. The states of the super agent is represented as $s_n = \{k_n, f_n, h_{1:n-1}, \varpi_n, d_n\}$. The vector $k_n$ contains real-time network indicators obtained from the analytics modules and \gls{RAG} system, while $f_n$ provides short-horizon forecasts such as predicted traffic load or packet-drop rates. The term $h_{1:n-1}$ reflects the historical record of previously invoked agents and their effects on the network. $\varpi_n$ captures the partial \gls{QoS} or intent-fulfillment status to indicate how closely the current system conditions align with the intended \gls{KPI} improvements. Finally, the drift and autonomous-trigger variable $d_n$ flags potential \gls{QoS} degradation or the absence of a new human intent. Also, $g_n = (m_n, \delta_n, z_n)$ denotes the goal state extracted from the operator intent, consisting of the target \gls{KPI} $m_n$, the desired improvement margin $\delta_n$, and the slice of interest $z_n$. The action $a_n=\{\mathrm{A_{ID}}(n),\theta_n\}$ represents the selection of an agent and its configuration parameters, while the reward function is based on (\ref{eq:pf}), which reinforces \gls{QoS} fulfillment and penalizes violations. Actions are defined at global scheduling epochs, but the super-agent acts only when events warrant it, making its action frequency non-periodic rather than per-\gls{TTI}. The generated trajectories $\mathcal{D}_{\text{offline}}=\{(s_n,g_n,a_n)\}$ form the dataset for policy learning. This dataset is generated using a hierarchical deep-Q-network that optimizes \eqref{eq:pf}, while \gls{HDM} is trained offline to imitate the resulting event-triggered orchestration policy.

Mamba is a neural network architecture in the state-space family that works like a modern recurrent neural network. It keeps a hidden state that updates at every step, but the update is selective and depends on the input, which helps the model focus on important tokens and ignore irrelevant ones \cite{8}. This selective update lets Mamba architecture handle long sequences with linear complexity. Decision Mamba applies this mechanism to reinforcement learning by treating states, actions, rewards, and goals as a sequence and predicting the next action in a stable and efficient way. \gls{HDM} extends this by adding a high-level module that selects sub-goals and a low-level module that produces actions, both using the same selective state-space design as presented in \cite{7} to manage long horizons and sparse rewards.

Our proposed \gls{HDM} is a hierarchical goal-conditioned structure built upon Mamba, and it is used for introducing decision intelligence to our super agent \cite{7}. This design accelerates convergence and reduces memory overhead for long-horizon orchestration tasks. Unlike \gls{DT} (one of our baselines) that depends on manually specified returns-to-go, \gls{HDM} learns to generate actions directly from goals \(g_t\) extracted from operator intents.

It consists of two hierarchical modules: a high-level \textit{meta-Mamba} that identifies a significant past action \(a_{n-\beta}\) contributing to goal progress, and a low-level \textit{control-Mamba} that predicts the next orchestration action \(a_n\) based on recent states, the goal, and the retrieved past action. Each input element: state, goal, and past action is encoded as latent embeddings and processed through Mamba’s continuous-time selective recurrence:
\begin{equation}
h_n = A(\Delta_n)h_{n-1} + B(\Delta_n)x_n, \quad y_n = C(\Delta_n)h_n + D x_n,
\end{equation}
where $A(\cdot), B(\cdot), C(\cdot), D(\cdot)$ are learnable matrices and $\Delta_t$ denotes the adaptive time step. The network implicitly models temporal ordering without explicit attention or positional encodings. Both Mamba modules are jointly trained to minimize the mean-squared error between predicted and target actions over $\mathcal{D}_{\text{offline}}$.

During inference, each validated intent is converted into a goal vector $g_n$, and the hierarchical Mamba modules collaboratively generate context-aware, goal-aligned actions. 

\textbf{2) Inter-slice agent:} In the proposed Agentic AI framework, the inter-slice agent manages how radio resources are allocated across different network slices. Inspired by \cite{2}, a \gls{DRL}-based inter-slice agent is designed. The observation space at global step $n$ is defined as $O_s(n) = [I_v, S_{mv}]$, where $I_v$ is the vector containing the QoS intents associated with the slice. For \gls{eMBB}, \gls{URLLC}, and \gls{BE} slices, we define $I_v$ as $I_v=[r_{embb}^{req}, \ell_{embb}^{req}, p_{embb}^{req}]$, $I_v=[r_{urllc}^{req}, \ell_{urllc}^{req}, p_{urllc}^{req}]$, and $I_v=[g_{be}^{req}, f_{be}^{req}]$, respectively. The slice metric vector $S_{mv}$ contains the corresponding performance metrics such as served long-term throughput, buffer occupancy, packet loss rate, and average buffer latency. 

We represent the action at decision step $n$ as a vector $A_n = [a_1, a_2, \ldots, a_S]$, where each component $a_s$ denotes the action factor for slice $s$ and takes a value in the interval $[-1, 1]$, consistent with the continuous outputs of the Gaussian distribution \cite{2}. The selected action vector $A_n$ is then translated into one of the $R_n$ discrete action choices through the following mapping function \cite{2}:
\begin{equation}
\text{index}(n) = \arg\min_{\text{option}} \left( d\left( R_{\text{comb}}^{\text{option}}, \left( R \frac{A_n + 1}{\sum_{i=1}^{S}(a_i + 1)} \right) \right) \right),
\label{eq:action_mapping}
\end{equation}
where $R_{\text{comb}}^{\text{option}}$ is one of the possible combinations of the \glspl{RBG}, $d$ calculates the Euclidean distance between the options available at $R_{\text{comb}}$ and \gls{RL} agent output $A_n$. Finally, the scheduling decision applied in the \gls{RL} environment (network system) is $R_{\text{comb}}^{\text{index}}$.

The reward function $\mathrm{RW}(n)$ utilizes slice-specific \gls{QoS} intents to evaluate the proximity of slice performance metrics to their intended targets. It comprises individual components corresponding to each slice type, defined as follows:  
\begin{align} \label{RW-func_main}
\text{RW}(n) = 
& \sum_{i \in \mathcal{S}_{1}} \text{R}_{1,i}(n)
+ \sum_{i \in \mathcal{S}_{2}} \text{R}_{2,i}(n)
+ \sum_{i \in \mathcal{S}_{3}} \text{R}_{3,i}(n),
\end{align}
where the $\text{R}_{\text{1},i}(n)$, $\text{R}_{\text{2},i}(n)$, and $\text{R}_{\text{3},i}(n)$ represent the reward for \gls{eMBB}, \gls{URLLC}, and \gls{BE} slice with index $i$ at step $n$.

The served throughput $\text{R}_{\text{embb}}^r(n)$, average buffer latency $\text{R}_{\text{embb}}^{\ell}(n)$, and the packet loss rate $\text{R}_{\text{embb}}^p(n)$ compose the eMBB slice reward function: 
\begin{equation} \label{embb_rw}
\text{R}_{\text{1},i}(n) = -\left( \text{R}_{\text{embb}}^r(n) + \text{R}_{\text{embb}}^{\ell}(n) + \text{R}_{\text{embb}}^p(n) \right).
\end{equation}

$\text{R}_{\text{embb}}^r(n)$, $\text{R}_{\text{embb}}^{\ell}(n)$, and $\text{R}_{\text{embb}}^p(n)$ are defined using the following equations.  
\begin{equation} \label{rate}
\text{R}_{\text{embb}}^r(n) =
\begin{cases}
w_{\text{embb}}^r \dfrac{r_{\text{embb}}^{\text{req}} - r_{\text{embb}}(n)}{r_{\text{embb}}^{\text{req}}}, & \text{if } r_{\text{embb}}(n) < r_{\text{embb}}^{\text{req}}. \\
0, & \text{if } r_{\text{embb}}(n) \geq r_{\text{embb}}^{\text{req}}.
\end{cases}
\end{equation}
\begin{equation} \label{latency}
\text{R}_{\text{embb}}^{\ell}(n) =
\begin{cases}
w_{\text{embb}}^{\ell} \dfrac{\ell_{\text{embb}}(n) - \ell_{\text{embb}}^{\text{req}}}{\ell_{\max} - \ell_{\text{embb}}^{\text{req}}}, & \text{if } \ell_{\text{embb}}(n) > \ell_{\text{embb}}^{\text{req}}. \\
0, & \text{if } \ell_{\text{embb}}(n) \leq \ell_{\text{embb}}^{\text{req}}.
\end{cases}
\end{equation}
\begin{equation} \label{pdr}
\text{R}_{\text{embb}}^p(n) =
\begin{cases}
w_{\text{embb}}^p \dfrac{p_{\text{embb}}(n) - p_{\text{embb}}^{\text{req}}}{1 - p_{\text{embb}}^{\text{req}}}, & \text{if } p_{\text{embb}}(n) > p_{\text{embb}}^{\text{req}}. \\
0, & \text{if } p_{\text{embb}}(n) \leq p_{\text{embb}}^{\text{req}}.
\end{cases}
\end{equation}

Here, $w_{\text{embb}}^r$ is the weight that defines the intent importance in relation to the other ones, $w_{\text{embb}}^{\ell}$ works as a weight for average buffer latency intent, and $w_{\text{embb}}^p$ is the weight factor for packet loss rate intent.

Similarly, we calculate $ \text{R}_{2,i}(n)$ and $\text{R}_{3,i}(n)$. More details on the inter-slice agent used in this work can be found in \cite{2}. 

\textbf{3) Intra-slice agent:}
In the proposed Agentic AI framework, each intra-slice agent allocates the \glspl{RBG} assigned by the inter-slice agent to the UEs in its slice. During each global scheduling step $n$, the intra-slice agent receives $R_s(n)$ \glspl{RBG} and performs $T_s = R_s(n)$ micro-steps indexed by $\kappa =\{ 1,\dots,T_s\}$, assigning one \gls{RBG} per micro-step. Within each slice $s \in \{1,\dots,S\}$, the micro-step state at $(n,\kappa)$ is denoted $s^{(s)}_{n,\kappa}$ and captures slice-specific features:
\begin{equation}
s^{(s)}_{n,\kappa} =
\begin{cases}
(\mathbf{q}, \mathbf{c}, \boldsymbol{\ell}, R_s(n)), & \text{URLLC},\\[3pt]
(\mathbf{q}, \mathbf{c}, \mathbf{u}, R_s(n)), & \text{eMBB/BE},
\end{cases}
\end{equation}
where $\mathbf{q}$, $\mathbf{c}$, $\boldsymbol{\ell}$, and $\mathbf{u}$ denote normalized queue occupancies, channel qualities, head-of-line delays, and service shares, respectively. The action 
$a^{(s)}_{n,\kappa} \in \{1,\dots,N_s\}$ selects the UE to receive the RBG at micro-step $\kappa$, updating queue, latency, and service-share statistics.

Each metric is normalized to $[0,1]$ as $\tilde{\ell}_j(n,\kappa)=\ell_j(n,\kappa)/\ell_{\max}$,
$\hat{b}_j(n,\kappa)=b_j(n,\kappa)/b_{\max}$, and
$u_j(n,\kappa)=\mathrm{RBG}_j(n,\kappa)/R_s(n)$. Here $\ell_j(n,\kappa)$ is the instantaneous head-of-line delay, 
$b_j(n,\kappa)$ is the number of bits served to UE $j$, 
$\mathrm{RBG}_j(n,\kappa)$ is the number of RBGs assigned to UE $j$ so far in step $n$, and $\bar{u}(n,\kappa)$ is the mean service share across all UEs in slice $s$. The intra-slice reward $r^{(s)}(n,\kappa)$, bounded within $[-1,1]$, is given by
\begin{equation}
r^{(s)}(n,\kappa)=
\begin{cases}
1-2\tilde{\ell}_j(n,\kappa), & s=\text{URLLC},\\[3pt]
2\hat{b}_j(n,\kappa)-1, & s=\text{eMBB},\\[3pt]
1-2|u_j(n,\kappa)-\bar{u}(n,\kappa)|, & s=\text{BE}.
\end{cases}
\end{equation}
This formulation enables the intra-slice \gls{DRL} agent to minimize latency, maximize throughput, and maintain fairness within each slice.

\begin{figure*}[!ht]
\centerline{\includegraphics[width=1\linewidth]{Fig2.eps}}
\caption{Performance comparison: (a) URLLC latency, (b) eMBB throughput, and (c) 5\textsuperscript{th} percentile throughput (proposed method vs. the baselines) (d) Self-healing property.} 
\vspace{-15pt}
\label{fig2}
\end{figure*}

\textbf{4) Inter-slice Self-healing Agent:} The inter-slice self-healing agent is modeled as an \gls{MDP}, where the state $s_n \in \mathcal{S}$ represents the observed network conditions relevant to corrective adaptation:$ s_n = [D_n,\, L_n,\, \bar{A}_n]$.
Here, $D_n = [\delta_n^{(1)},\dots,\delta_n^{(m)}]$ denoting slice-level QoS deviations, 
$\delta_t^{(i)} = \mathrm{QoS}_{\mathrm{desired}}^{(i)} - \mathrm{QoS}_{\mathrm{actual}}^{(i)}$, 
$L_n$ the instantaneous system load, and $\bar{A}_t$ the current inter-slice \gls{RBG} allocation.  

The action $a_n = [\Delta w_1, \ldots, \Delta w_m]$ applies bounded weight adjustments $\Delta w_i \in [-\Delta_{\max}, \Delta_{\max}]$ to modify slice priorities in the inter-slice scheduler.  
The reward measures the degree of \gls{SLA} compliance and is slice-specific. It is defined as: 
$\mathcal{R}_n=\dfrac{T_{\mathrm{e}}(n)}{T_{\mathrm{e}}^\star}$ (eMBB), $\mathcal{R}_n=\dfrac{D_{\mathrm{u}}^\star}{D_{\mathrm{u}}(n)}$ (URLLC), and $\mathcal{R}_n=\dfrac{Q^{(5\%)}_{\mathrm{b}}(n)}{(Q^{(5\%)}_{\mathrm{b}})^\star}$ (BE). Here, $D_{\mathrm{u}}(n)$ is the measured delay and $D_{\mathrm{u}}^\star$ is the maximum allowable delay defined by the \gls{QoS} requirement. Also, $T_{\mathrm{e}}(n)$ denotes the instantaneous throughput and $T_{\mathrm{e}}^\star$ represents the \gls{QoS}-defined target throughput. Lastly, \( Q^{(5\%)}_{\mathrm{b}}(n)\) denotes the \( 5^{\mathrm{th}} \) percentile throughput at time \(n\), and \( \left(Q^{(5\%)}_{\mathrm{b}}\right)^\star \) denotes the target \( 5^{\mathrm{th}}\) percentile throughput specified to ensure that even the lowest-rate users achieve a minimum acceptable level of service.

In our proposed framework, agents are co-located control modules within the same control plane, hence communicate through structured function calls. All the agent actions and outcomes are centrally observed and recorded by the super agent. The super agent maintains a global view of system state, active allocations, and intent constraints to avoid conflicts.

\section{Performance Evaluation}
\label{s5}

\subsection{Simulation Setup}

The proposed framework is evaluated using a mmWave multi-cell \gls{5G NR} simulation aligned with the system model. A seven-cell hexagonal layout with three sectors per site is used, where the central sector is assessed while the surrounding sectors create inter-cell interference. The system operates at 30 GHz with 100 MHz bandwidth and 60 kHz subcarrier spacing ($\mu=2$). Each base station uses a 64-element uniform linear array, and user terminals use 4-element arrays \cite{10}. Propagation follows the 3GPP clustered delay line fading model with distance-based path loss and log-normal shadowing \cite{11}. Three slice types are considered: \gls{eMBB}, \gls{URLLC}, and \gls{BE} with Poisson traffic arrivals and \gls{BE} users activated/deactivated over time. Performance constraints such as \gls{URLLC} latency being $\leq$ 2ms, \gls{eMBB} aggregate throughput $\geq$150 Mbps are enforced, which also guide reward design and the self-healing behavior. The simulation is implemented via MATLAB–Python co-simulation: MATLAB handles physical layer, channel realization, mobility, and RBG-level throughput, while the inter-slice, intra-slice \gls{RL} agents and the \gls{HDM} controller are implemented using Python.

\subsection{Simulation Results}

To ensure a fair evaluation, we compare the proposed \gls{HDM}-guided Agentic AI framework against three baselines. The first baseline, Agentic AI with Offline Decision Transformer. This particular baseline preserves the same Agentic architecture but replaces the \gls{HDM} controller with an offline \gls{DT} model. The second baseline is \gls{HRL}, which also maintains the same design, but there is no Agentic control for self-healing actions. Finally, a non-ML baseline with fixed rule-based scheduling and static slice priorities is also considered as a traditional baseline. 

\gls{URLLC} latency grows with the number of \glspl{UE} due to increased queueing (Fig.~\ref{fig2}a); however, the proposed \gls{HDM}-enabled Agentic AI maintains significantly lower latency. The \gls{eMBB} aggregate throughput rises with increasing traffic load (Fig.~\ref{fig2}b). Across all load levels, the proposed method achieves approximately $3\%$, $11\%$, and $16\%$ higher throughput relative to \gls{DT}, \gls{HRL}, and non-ML scheduling, respectively. The 5th-percentile throughput, which represents cell-edge users, degrades at high load for all schemes (Fig.~\ref{fig2}c), yet the proposed framework delivers consistently higher fairness, improving by $8\%$, $20\%$, and $34\%$ over the same baselines. The delay is reduced by $19\%$ compared to \gls{DT}, $28\%$ to \gls{HRL}, and $43\%$ to the rule-based approach at peak traffic. These gains stem from the linear-time selective recurrence of \gls{HDM}, which lowers control latency, reduces memory overhead, and preserves long-range dynamics such as past resource allocations and their impact. Therefore, the system can achieve near–real-time responsiveness even under heavy load. Moreover, when \gls{QoS} drift occurs, the \gls{HDM}-driven Agentic AI autonomously restores the degrading \gls{KPI}. Fig.~\ref{fig2}d illustrates such a self-healing event for \gls{URLLC} latency, where the proposed method recovers rapidly, outperforming \gls{DT} and \gls{HRL}, while the rule-based scheduler fails to respond. All the results are obtained from $10$ independent simulation runs, each initialized with different random seeds, UE placements, traffic arrivals, and channel realizations.

\section{Conclusions}

In this work, we introduced an Agentic AI framework for 6G \gls{RAN} slicing supported by an \gls{HDM} controller, where an \gls{LLM}-based super agent interprets operator intents and coordinates resource allocation across slices. \gls{HDM} enables faster and more context-aware orchestration than transformer-based and reward-driven baselines. Simulation results show consistent improvements across key \glspl{KPI}, including higher \gls{eMBB} throughput, better cell-edge performance, and lower \gls{URLLC} latency, outperforming the baselines. These gains are enabled by the integration of H-RAG for up-to-date contextual knowledge, combined with coordinated agent interactions guided by the Decision Mamba controller.

\label{s6}

\section*{Acknowledgement}

This work has been supported by MITACS, Ericsson Canada, and the NSERC Canada Research Chairs program.

\bibliographystyle{IEEEtran}
\bibliography{reference.bib}
\end{document}

%% file: glossary.tex
\newacronym{RAN}{RAN}{Radio Access Network}
\newacronym{RRS}{RRS}{Radio Resource Scheduler}
\newacronym{LLM}{LLM}{Large Language Model}
\newacronym{SLA}{SLA}{Service Level Agreement}
\newacronym{5G}{5G}{Fifth-Generation}
\newacronym{HRL}{HRL}{Hierarchical RL}
\newacronym{QoS}{QoS}{Quality-of-Service}
\newacronym{RAG}{RAG}{Retrieval Augmented Generation}
\newacronym{KPI}{KPI}{Key Performance Indicator}
\newacronym{RL}{RL}{Reinforcement Learning}
\newacronym{DT}{DT}{Decision Transformer}
\newacronym{HDM}{HDM}{Hierarchical Decision Mamba}
\newacronym{RAT}{RAT}{Radio Access Technology}
\newacronym{NR}{NR}{New Radio}
\newacronym{LTE}{LTE}{Long-Term Evolution}
\newacronym{MIMO}{MIMO}{Multiple Input Multiple Output}
\newacronym{BS}{BS}{Base Station}
\newacronym{UE}{UE}{User Equipment}
\newacronym{RBG}{RBG}{Resource Block Group}
\newacronym{TDD}{TDD}{Time Division Duplexing}
\newacronym{SE}{SE}{Spectral efficiency}
\newacronym{H-RAG}{H-RAG}{Hybrid RAG}
\newacronym{LoS}{LoS}{Line-of-Sight}
\newacronym{URLLC}{URLLC}{Ultra-Reliable Low-Latency Communication}
\newacronym{MDP}{MDP}{Markov Decision Process}
\newacronym{eMBB}{eMBB}{Enhanced Mobile Broadband}
\newacronym{BE}{BE}{Best Effort}
\newacronym{TTI}{TTI}{Transmission Time Interval}
\newacronym{RSRP}{RSRP}{Reference Signal Received Power}
\newacronym{DRL}{DRL}{Deep RL}
\newacronym{PRB}{PRB}{Physical Resource Block}
\newacronym{5G NR}{5G NR}{5G New Radio}